# Radiation hardness of ultrabroadband spintronic terahertz emitters: en-route to a space-qualified terahertz time-domain gas spectrometer

O. Gueckstock[1,2], N. Stojanovic[3], Yookyung Ha[3], T. Hagelschuer[3], A. Denker[4], G. Kourkafas[4], T. S. Seifert[1,5], T. Kampfrath[1,2], M. Gensch[3,6,*]

1. Department of Physics, Freie Universität Berlin, Arnimallee 14, 14195 Berlin, Germany
2. Department of Physical Chemistry, Fritz Haber Institute of the Max Planck Society, Faradayweg 4-6, 14195 Berlin, Germany
3. Institute of Optical Sensor Systems, DLR (German Aerospace Center), Rutherfordstr. 2, 12489 Berlin, Germany
4. Helmholtz-Zentrum Berlin für Materialien und Energie, Albert-Einstein-Str. 15, 12489 Berlin, Germany
5. TeraSpinTec GmbH, Mühlenstraße 12, 14959 Trebbin, Germany
6. Institute of Optics and Atomic Physics, Technische Universität Berlin, Strasse des 17. Juni 135, 10623 Berlin, Germany

* E-Mail: michael.gensch@dlr.de

## Abstract

The radiation hardness of ultrabroadband, spintronic terahertz emitters against $\gamma$ and proton irradiation is investigated. We find that irradiation doses equivalent to those experienced by a space instrument en-route to and operated on Mars have a minor effect on the performance of the emitter. In particular, the ultrawide emission spectrum 0.1-30 THz, which covers a large part of the vibrational fingerprint region, remains unchanged. These results make this emitter type highly interesting as essential building block for broad-band gas sensors based on terahertz time-domain spectroscopy for future space missions.

## Results

The beginning of the 21st century has seen the emergence of robust, turnkey femtosecond laser systems and associated time-domain spectroscopy (TDS) techniques [1]. Since then, these techniques have revolutionized laboratory spectroscopy by, e.g., providing straightforward access to the low terahertz (THz) frequency range (~0.1-3 THz) and increasingly also the far-infrared and mid-infrared spectral ranges. The THz and infrared spectral windows are of high relevance for the exploration of space [2, 3, 4] because many optical resonances of planetary materials are located in this frequency range.

So far, there have been no reports of a THz-TDS instrument operating in outer space, and THz and infrared spectroscopy instruments still rely on the Fourier-transform infrared (FTIR) principle (see, e.g., [3, 5] and references therein) or grating spectrometers [6] in combination with direct, often cryogenically cooled detectors. Because THz-TDS is based on stroboscopic sampling of coherent THz electromagnetic waveforms, the achievable bandwidth is ultimately governed by the laser pulse duration, which is required to be 100 fs or smaller to reach the THz range. One reason for the lack of THz-TDS space instruments has been the absence of space-qualified femtosecond lasers, which are the prerequisite for TDS [1]. This situation has fundamentally changed because the space qualification of compact fiber lasers has been verified over the past decade [7, 8, 9].

The other apparent drawback of THz-TDS with respect to FTIR instruments has been the limited bandwidth originating largely from THz wave attenuation or absorption in classical THz emitters like ZnTe, LNbO and GaAs [1, 10]. To this end, a promising emitter class, spintronic THz emitters (STEs), has emerged [11]. As STEs are made of metal, they do not exhibit bandwidth limitations due to Reststrahlen bands or absorption and provide a broad spectrum ranging from 0.1 THz to 30 THz [11, 12].

These two recent developments, space-qualified femtosecond lasers and broadband THz emitters, make THz-TDS systems a true alternative to FTIR or grating instruments in space research, since they can be extremely compact, are possibly fully chip-integrable and avoid the necessity of cryogenic detector systems.

In this letter, we first show by example how powerful THz time-domain spectroscopy can be used for the identification of spectral fingerprints of gases. Second, we test the radiation hardness of spintronic THz emitters (TeraSpinTec GmbH) by radiation-hardness tests with doses of low-energy $\gamma$-radiation and high-energy protons equivalent to those utilized to verify instrument space qualification for a mission to Mars [13].

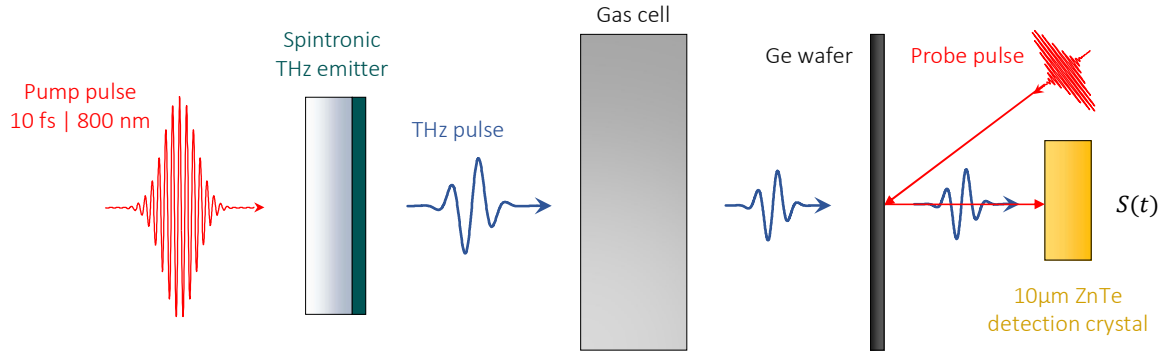

**FIG. 1. Ultrabroadband THz transmission spectroscopy.** A femtosecond laser pulse excites a spintronic THz emitter generating a broadband THz pulse with frequencies ranging from 1-30 THz. The emitted pulse traverses a cell with a sample gas. The resulting THz electric field $E(t)$ is measured by electro-optic detection (EOS) in a ZnTe(110) crystal versus time $t$. For referencing, an empty gas cell is used.

To generate ultrabroadband THz pulses, we use 10 fs long near-infrared pulses (800 nm, 80 MHz, 2.5 nJ) from a femtosecond Ti:Sapphire laser oscillator and excite a metallic spintronic heterostructure W|CoFeB|Pt deposited on a sapphire substrate (0001, 500µm-thick) consisting of a ferromagnetic CoFeB and two non-magnetic W and Pt layers [11]. Upon excitation by the laser pulse a spin voltage and, thus, a longitudinal spin current is generated [14] in the vicinity of the W/CoFeB and CoFeB/Pt interfaces and is converted into two constructively superimposed charge currents by the inverse spin Hall effect in W and Pt. Due to the pulsed excitation, the charge current is time-dependent and acts as a source of an electromagnetic pulse with frequencies in the THz regime [11, 15].

The transient electric field of the emitted THz pulse is characterized by electro-optic sampling (EOS) [16, 17] in a 10 µm thick ZnTe(110) crystal, providing electro-optic signals $S(t)$ versus time $t$ with a broad detection bandwidth of beyond 30 THz (see Fig. 1). The THz waveforms are sampled over a time window of 8-10 ps providing a frequency resolution of 100-125 GHz (3.3-4.2 cm$^{-1}$). Before EOS, the THz pulse traverses a sample gas in a cell. Measurements of $S(t)$ are taken for the whole THz setup purged with ambient air (laboratory air) or with dry air.

We first address the spectroscopic power of our setup and subsequently study the radiation-hardness of the STE.

Fig. 2a shows typical THz electro-optic signals $S(t)$ from the STE under ambient-air and dry-air conditions. The corresponding Fourier amplitude spectra $\tilde{S}(\omega)$ (Fig. 2b) have sufficient intensity over more than 50% of the molecular fingerprint region between 400 and 1000 cm$^{-1}$ [18] as well as the lower THz frequency range between 10 and 400 cm$^{-1}$. This bandwidth provides access to many characteristic eigenfrequencies of the rotational-vibrational spectrum of molecules.

The spectra taken in dry air (Fig. 2b) contain primarily features arising from components of the THz-TDS setup themselves. A zero in the electro-optic coefficient of the ZnTe detector gives rise to the dip observed at 5 THz [16], while the broad features at around 13 THz, 17 THz and 19 THz originate from features in the refractive index of the sapphire substrate of the STE

(gray shaded area) [19, 20]. The dominant feature at 10 THz originates from a two-phonon process in the Ge wafer [21] that is utilized to remove the near-infrared radiation after the STE. Note that the two-phonon absorption in Ge is relatively broad band and influences the spectra from 8 to 13 THz. The spectra taken under ambient conditions show, in addition, a rich set of absorption lines from the rotational eigenfrequencies of water vapor [22] at frequencies below 13 THz and of one absorption line from the bending mode of $CO_2$ at around 20 THz.

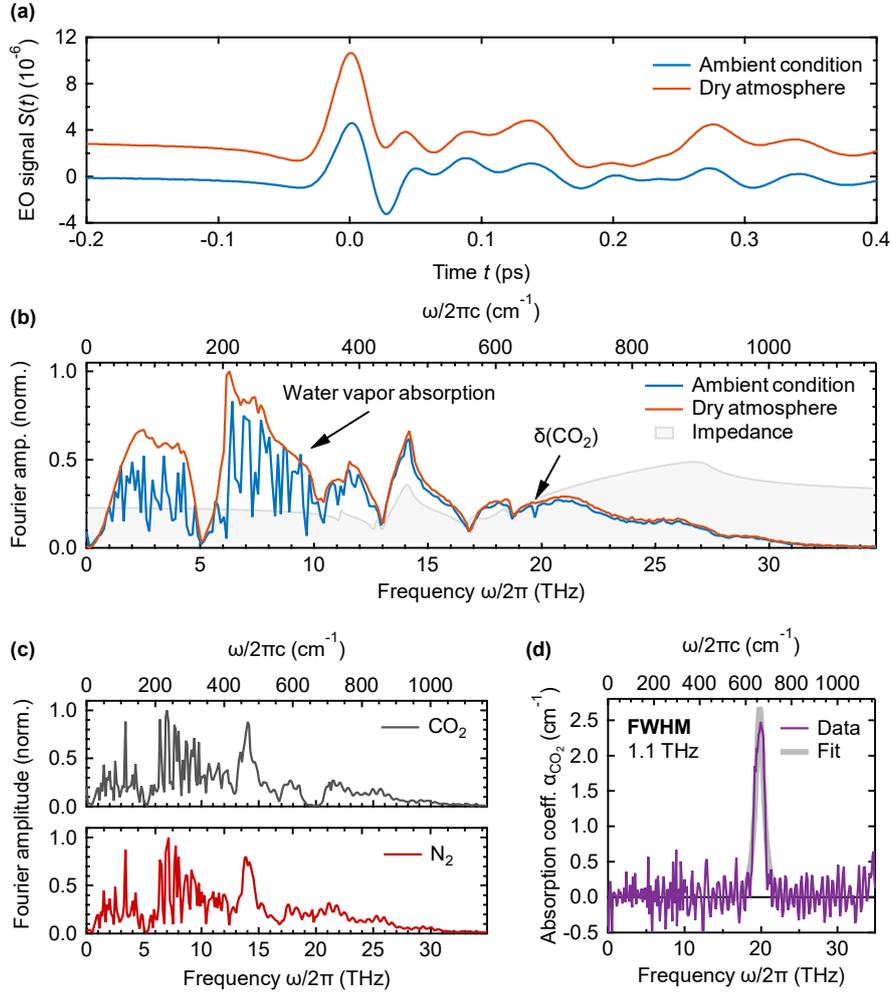

**FIG. 2. THz spectroscopy for identification of water vapor and $CO_2$.** (a) Broadband THz emission waveforms from a spintronic THz emitter (STE) under ambient conditions (blue) and in a dry-air atmosphere (orange). Waveforms are vertically offset for clarity. (b) Normalized amplitude spectra of the waveforms in panel (a). The gray shaded area shows the impedance of the STE. (c) THz transmission spectra from an STE THz pulse following transmission of a gas cell filled with 50 mbar of carbon dioxide $CO_2$ (top panel) and 50 mbar of molecular nitrogen $N_2$ (bottom panel). Note the dip around 20 THz in the $CO_2$ transmission spectrum. (d) $CO_2$ absorption coefficient $\alpha_{CO_2}$ as extracted from panel (c).

To verify the application for gas detection, further measurements were performed on $CO_2$ and $N_2$ in the gas cell of the set-up (see Fig. S1). The results of these measurements are shown in Fig. 2c. As $N_2$ has no infrared-active modes, the measurement only exhibits residual water absorption bands and Fabry-Pérot-type interference features from THz-pulse echoes in the gas cell and probe-pulse echoes in the detection crystal. In contrast, the spectrum of $CO_2$ exhibits a strong absorption line due to two degenerate bending vibrations at around 20 THz. The absorption coefficient of $CO_2$ is determined by $\alpha_{CO_2} = -(1/d)\ln|\tilde{S}_{CO_2}(\omega)/\tilde{S}_{N_2}(\omega)|$ from the data in Fig. 2d, where $d$ denotes the length of the gas cell. The determined bandwidth of the absorption line of 1.1 THz FWHM (Lorentzian fit in grey) is governed by the limited range of the time delay $t$ utilized in the experiment.

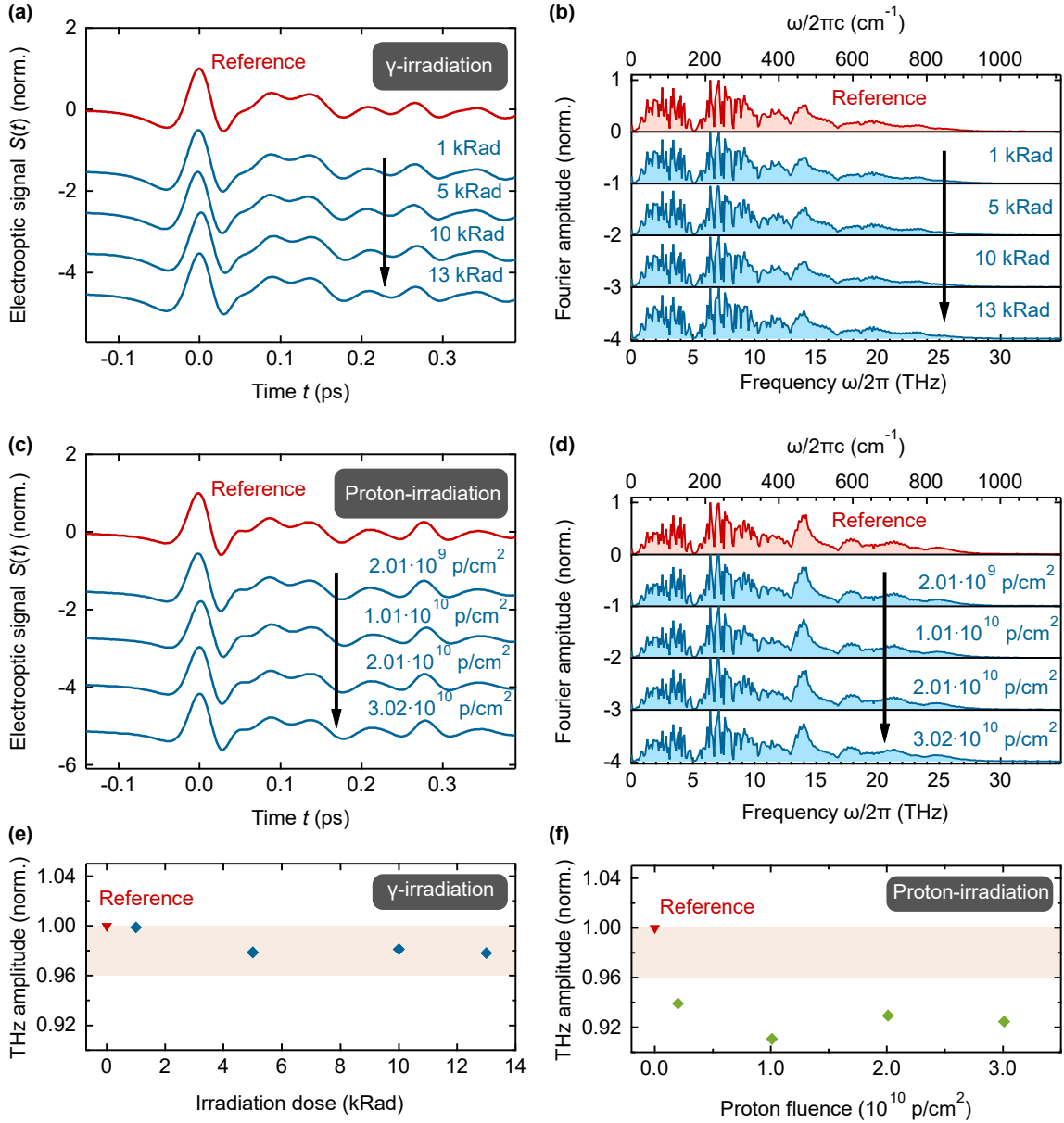

**FIG. 3. STE radiation-hardness. (a)** Time-domain waveform and **(b)** frequency-domain spectrum for different irradiation levels of γ radiation from a Co60 source. All waveforms are vertically offset for clarity. **(c)** Time-domain waveform and **(d)** frequency-domain spectrum for different irradiation levels of high energy 68 MeV protons from a cyclotron accelerator. Dependence of the THz signal amplitude on the irradiation dose with **(e)** γ radiation and **(f)** high-energy protons. The light orange bar in the background of panels (e,f) highlights the variation of 4% of the emitted THz pulse amplitude from STEs out of the same growth batch and systematic uncertainties. Note that the proton flux was constant at roughly $1.1 \times 10^7$ p/cm²s.

The working principle of the STE enables unprecedented, broadband THz emission and hence broadband THz spectroscopy. However, it is based on very thin films and ferromagnetic order, and spin transport and its conversion into an in-plane charge current depend sensitively on the film interfaces [23]. It is, thus, not a priori clear whether the STE is robust under the various kinds of radiation found in space. More precisely, the efficiency of the THz emission depends sensitively on the amplitude of the laser-induced spin voltage in CoFeB because it is proportional to the time-derivative of the quenched magnetization [24]. Second, spin transport across the two CoFeB interfaces and, third, the spin-to-charge current conversion determine the efficiency of the emitted THz pulse. In fact, the static magnetic and electric transport properties of a variety of materials have been shown to be affected by irradiation with γ photons, protons and other high-energy particles [25, 26]. Therefore, it is of utmost importance for applications of STEs in space research to verify in what way their emission efficiency is affected by the inevitable irradiation doses received during the transport through space and operation in extraterrestrial environments.

For that purpose, identical STEs were exposed to controlled radiation doses of low-energy $\gamma$ radiation from a Co60 source and 68 MeV protons from a cyclotron accelerator [27] (see Figure S3 and S4 for details). The highest $\gamma$ radiation level of 13 kRad and proton fluence of $3 \times 10^{10}$ p/cm$^2$ were chosen as they corresponds to those expected for a journey of the STE from Earth to Mars and for subsequent operation on one of the Martian moons shielded by an Aluminum absorber of ~2 mm thickness including a radiation design margin factor of 1.25 [13]. One STE was kept unirradiated to serve as a reference. Thereafter, THz-emission studies were performed to analyze the effects of the irradiation on amplitudes and the spectral content.

The results of these benchmark experiments are shown in Fig. 3. The signal waveforms of the emitted THz radiation (Fig. 3a,c), their amplitudes (Fig. 3e,f) and overall amplitude spectra (Fig. 3b,d and Fig. S2) is largely unaffected (±4% amplitude changes) by irradiation with $\gamma$ rays, as can be seen in Fig. 3e. This behavior is in contrast to the exposure with protons where a moderate reduction of the THz amplitude by 8% occurs at low radiation doses. Importantly, after this initial drop, the THz signal amplitude remains constant for all doses up to 15 kRad (see Fig. 3f).

The spectral content of the emitted THz pulses is largely unaffected by irradiation with $\gamma$ radiation and protons. This finding indicates that the dynamics of the driving spin voltage as well as the subsequent spin-to-charge current conversion remain unchanged on femtosecond timescales. Regarding the observed minor drop of the emitted THz amplitude upon irradiation with protons, earlier work indicates that the efficiency of the THz emission process depends strongly on the interfaces of the spintronic heterostructure [23]. Thus, one could speculate that the proton irradiation at low dose rates irreversibly modifies the interfaces. The important fact that this process does not gradually continue with increasing dose rates cannot easily be understood at this stage and should be investigated in future experiments.

In any case, our-radiation hardness experiments clearly show that STEs consisting of metallic nanolayers of CoFeB, W and Pt fulfill the $\gamma$ and proton radiation-hardness criteria for a mission to Mars.

The presented extension of the bandwidth of THz-TDS to 0.1-30 THz using STEs is a major step toward application as compact gas sensors. Besides the bandwidth, the achievable frequency resolution is an important parameter. In the measurements presented here, the frequency resolution was limited to 0.1 THz (3.33 cm$^{-1}$) by the length of the optical delay line of 10 ps (path length of 3 mm). A resolution of 1 GHz (0.033 cm$^{-1}$) requires a path length of only 30 cm and can be easily realized. To achieve higher resolutions in gas spectroscopy down to the 100 MHz level (0.0033 cm$^{-1}$), Constrained Reconstruction Super Resolution (CRSR) algorithms have been recently proposed and demonstrated [28]. Thereby, a compact, space-qualified THz-TDS based gas sensor can be envisioned spanning a bandwidth of few decades in the THz frequency range and a frequency resolution in the sub-GHz regime.

## Acknowledgements

The authors thank Genaro Bierhance (FU Berlin) and Dr. Sergey Pavlov (DLR Institute of Optical Sensorsystems) for fruitful discussions. O.G., T.SS. and T.K. acknowledge funding by the Deutsche Forschungsgemeinschaft (DFG, German Research Foundation) through the Collaborative Research Center SFB TRR 227 "Ultrafast spin dynamics" (Project ID 328545488, projects B02 and A05).Y.K.H., N.S. and M.G acknowledge funding by the Deutsche Forschungsgemeinschaft (DFG, German Research Foundation) through the priority program SPP2314 INTEREST (project ITISA, project ID GE 3288 2-1).